\documentclass{article}
\usepackage{spconf,amsmath,graphicx}
\usepackage{multirow}
\usepackage{footnote}
\usepackage{hyperref}
\usepackage{subfigure}
\usepackage{amssymb}
\usepackage{booktabs}

\title{DWFormer: Dynamic Window transFormer for Speech Emotion Recognition}
\name{Shuaiqi Chen$^1$, Xiaofen Xing$^1$\sthanks{Corresponding Author. xfxing@scut.edu.cn}, Weibin Zhang$^2$, Weidong Chen$^1$, Xiangmin Xu$^3$ $^4$}
\address{
  $^1$School of Electronic and Information Engineering, South China University of Technology, China \\
  $^2$ VoiceAI Technologies, Shenzhen, China
  $^3$ Pazhou Laboratory, China \\
  $^4$ School of Future Technology, South China University of Technology, China
}
\begin{document}
%
\maketitle
\begin{abstract}
Speech emotion recognition is crucial to human-computer interaction. The temporal regions that represent different emotions scatter in different parts of the speech locally. Moreover, the temporal scales of important information may vary over a large range within and across speech segments. Although transformer-based models have made progress in this field, the existing models could not precisely locate important regions at different temporal scales. To address the issue, we propose Dynamic Window transFormer (DWFormer), a new architecture that leverages temporal importance by dynamically splitting samples into windows. Self-attention mechanism is applied within windows for capturing temporal important information locally in a fine-grained way. Cross-window information interaction is also taken into account for global communication. DWFormer is evaluated on both the IEMOCAP and the MELD datasets. Experimental results show that the proposed model achieves better performance than the previous state-of-the-art methods.
\end{abstract}
\begin{keywords}
speech emotion recognition, transformer, speech signal processing
\end{keywords}
\section{Introduction}
\label{sec:intro}

Speech Emotion Recognition (SER) is the key to human-computer interaction. To make human-computer interaction more natural, it is essential for machines to precisely capture emotions and respond in an appropriate manner.

SER has been studied for decades. In recent years, transformer-based models have fostered huge improvement in  SER field \cite{chen2022key, chen22_interspeech, wang2021novel,wang2021learning}. The vanilla transformer \cite{vaswani2017attention} is outstanding in modeling long-range dependencies in speech sequences. However, its core mechanism, global self-attention mechanism, is vulnerable to noise and may not be able to focus on the same areas as the location of the emotion \cite{chen2022key}. This limits the effectiveness of the transformer model. Sound events that prominently represent emotions, such as changes in intonation and speed, laughs and sighs, are located in local regions. Furthermore, the scales of important information are varied over a large range within and across speech segments (see Fig. 1). \cite{chen22_interspeech} applies local window attention mechanism to enable models to focus more on local changes.  However, immutable window lengths limit these models to capture sentiment information that varies with  different temporal scales. 
\begin{figure}[t]
    \centering
    \includegraphics[width = \linewidth]{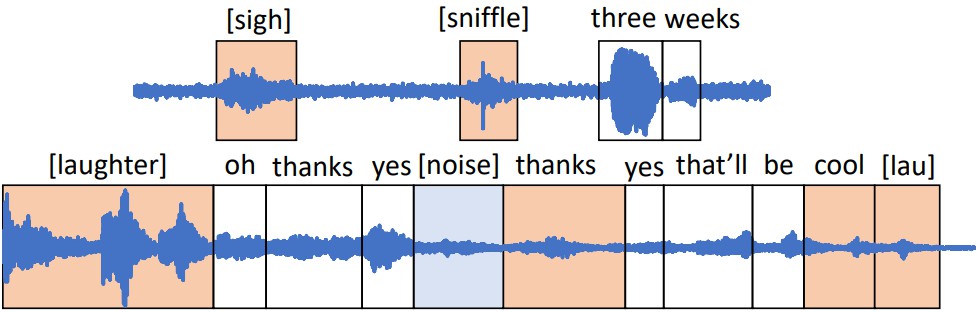}
    \caption{Two examples are selected from IEMOCAP\cite{busso2008iemocap}. [lau] represents laughter. Important sound events that indicates different emotions, such as laughter, sigh, sniffle and positive semantics etc., exist in local regions of speech and their duration varies.}
    \label{fig:my_label3}
\end{figure}
\begin{figure*}[t]
    \centering
    \includegraphics[width = \linewidth]{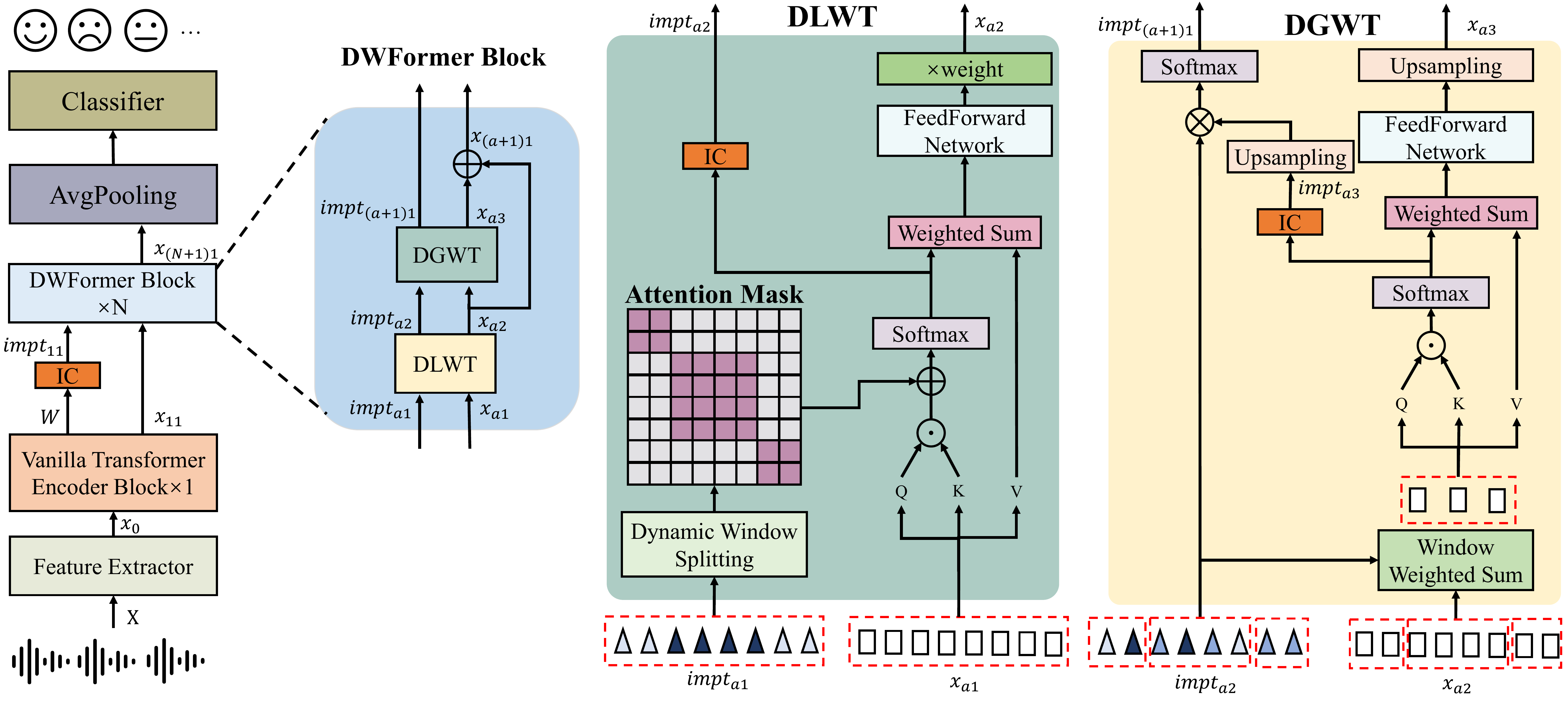}
    \caption{Model architecture of DWFormer. For simplicity, the residual connection and layer normalization are not plotted in the figure. IC represents Importance Calculation module (detailed in Fig. 3). The triangular sequence $impt$ means the important weights. The deeper the color, the more important the token is. The rectangle sequence $x$ represents feature map.}
    \label{fig:my_label3}
\end{figure*}

In computer vision field, dynamic designs \cite{ren2022beyond,NEURIPS2020_1963bd51,8907873} allow models to have a flexible perceptual field so that different shapes of targets can be captured. In SER field, \cite{liu2022atda, liu2020speech, liu2020temporal} propose local-global architectures to capture important temporal information. Inspired by them, a new architecture named Dynamic Window transFormer (DWFormer) is proposed to solve the aforementioned problem. The core of the proposed architecture, the DWFormer block, is composed of a Dynamic Local Window Transformer (DLWT) module and a Dynamic Global Window Transformer (DGWT) module. DLWT dynamically divides the input feature into several scales of windows and captures local important information in each window. DGWT remeasures the importance between windows after DLWT. The combination of DLWT and DGWT helps the model discover task-relevant regions.
The main contributions of this paper are as follows:

(\romannumeral1) A new architecture, named Dynamic Window transFormer (DWFormer), is proposed to provide insights into the problem of capturing important temporal information of variable lengths for SER. 

(\romannumeral2) We evaluate DWFormer on both the IEMOCAP and MELD \cite{poria2018meld} datasets and demonstrate that DWFormer substantially outperforms the vanilla transformer and the fixed window transformer. 
Besides, DWFormer achieves comparable results to the conventional studies and the state-of-the-art approaches. The code will be published.\footnote{https://github.com/scutcsq/DWFormer}
\section{Methodology}
\label{sec:format}
The architecture of DWFormer is shown in Fig. 2. The core of the model architecture is the DWFormer block, which is made up of a Dynamic Local Window Transformer module and another Dynamic Global Window Transformer module. The components of the model are introduced below.

The input audio signal is first fed into the feature extractor to extract the features $x_0\in \mathbb{R}^{T \times D}$, where $T$ represents the number of feature tokens, $D$ represents the feature dimension. Then $x_0$ is passed through a vanilla transformer encoder layer. The outputs of the encoder layer consists of the hidden feature $x_{11}$ and attention weights $W \in \mathbb{R} ^{H \times T \times T}$ where $H$ represents the number of heads.  $W$ is sent into the Importance Calculation module to obtain an temporal importance estimation which is necessary for the $1$st DWFormer block. 

\subsection{Importance Calculation Module}
 The Importance Calculation (IC) module is proposed to measure the importance of token. Inspired by \cite{chen2022key}, IC module utilize the attention weights obtained from transformer for calculation. The process is shown in Fig. 3, which is described as:
\begin{equation}
    impt = Softmax[\sum_1^{T_1} (\frac{1}{H} \sum_{s=1}^H aw_s)]
\end{equation}
where $aw_s$ represents attention weight from $s$-head,  $T_1$ is the row length of the averaged matrix $aw_{avg}$. Softmax function is used for normalization.
\begin{figure}[ht]
    \centering
    \includegraphics[width=\linewidth]{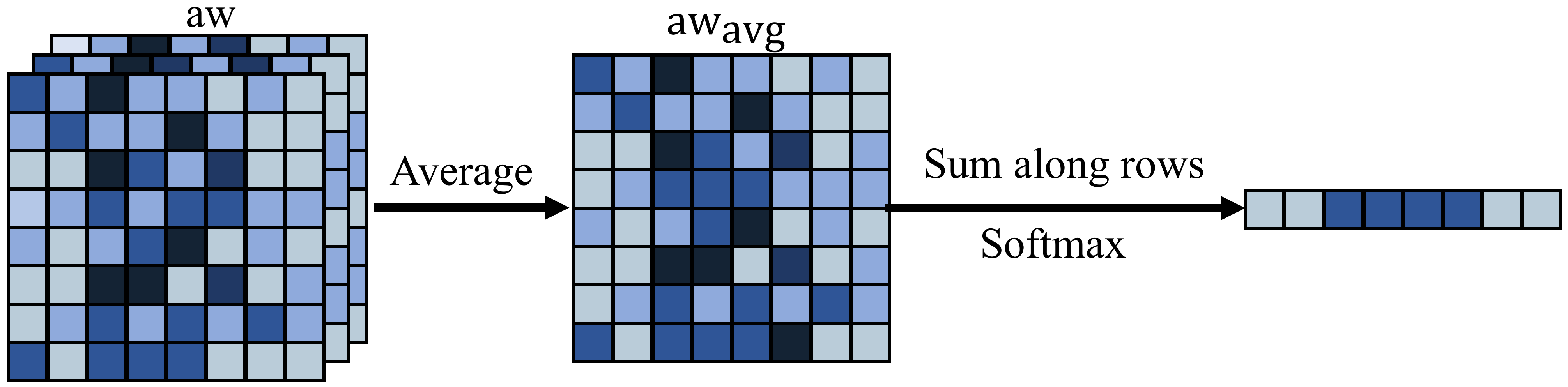}
    \caption{The IC module calculates the importance from the attention weights.}
    \label{fig:my_label}
\end{figure}

The importance score of each token $impt_{11}$ obtained by IC module, together with the hidden feature $ x_{11}$ are then transferred to N stacked DWFormer blocks for further evaluation.

\subsection{DWFormer Block}


\subsubsection{Dynamic Local Window Transformer Module}
The Dynamic Local Window Transformer (DLWT) module dynamically partitions regions for input feature and captures important information through local relationship modeling. The procedure is elaborated as bellows.

Firstly, utilizing \textbf{dynamic window splitting}(DWS) operation, feature tokens are dynamically split into unequal-length windows according to their importance values obtained from the IC component. As shown in Fig. 4, based on the importance scores calculated from the former block, tokens with importance scores above/below the threshold are grouped chronologically into several strong/weak emotional correlation windows. The threshold is set to the median of all the importance values. $A$ strong emotional correlation windows and $B$ strong emotional correlation windows are obtained from $x_{a1}$.

  To process data in batches, the window division results are implemented by attention mask mechanism:
\begin{align}
    \begin{split}
        M_{ij} = \left \{
    \begin{array}{lr}
    0, &  (b_{w_{k}} \leq i \leq 
     e_{w_{k}}, b_{w_{k}} \leq j \leq e_{w_{k}}\\ & k = 1,...,A+B) \\
    - \infty,     & otherwise
    \end{array}
    \right .
    \end{split}
\end{align}
where $M_{ij}$ is the value of $i$th row and $j$th column of the attention mask $M \in \mathbb{R}^{T \times T}$.  $b_{w_{k}}$ and $e_{w_{k}}$ are the begin and the end indexes of the row and column of the $k$th window.

\begin{figure}[ht]
    \centering
    \includegraphics[width = \linewidth]{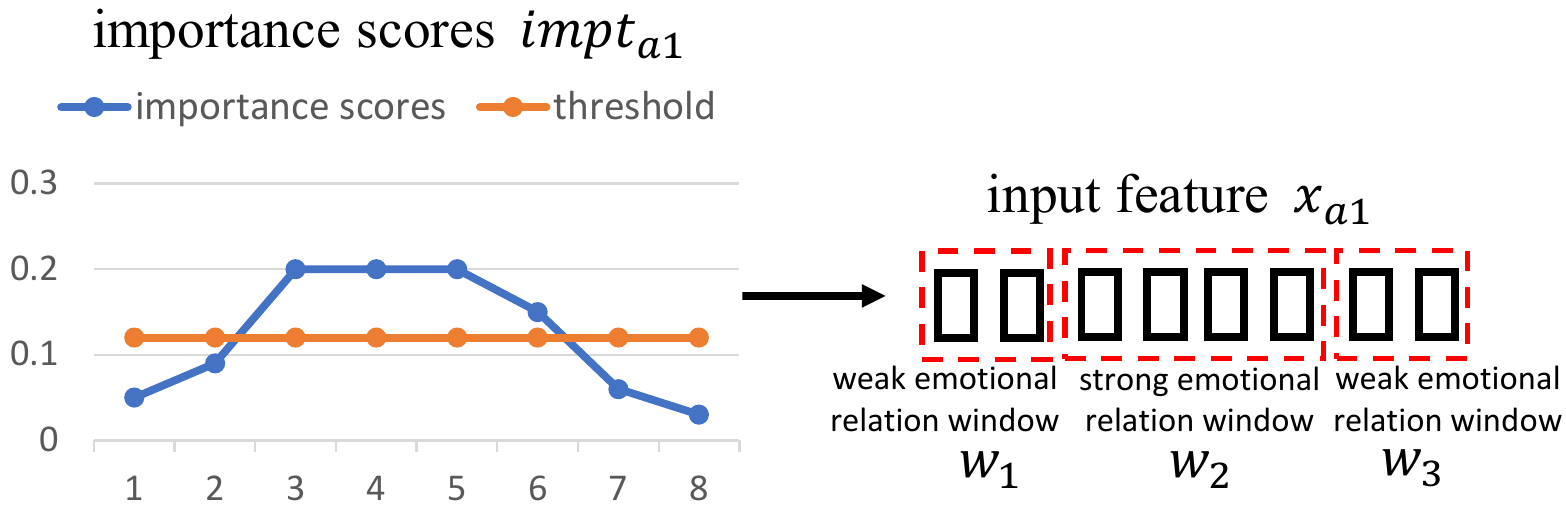}
    \caption{The operation of dynamic window splitting.}
    \label{fig:my_label3}
\end{figure}

 Then, each window passes through a transformer encoder for intra-window information communication, which is defined as:
 

\begin{equation}
    DLWT(x_{a1}) = FFN(Softmax(\frac{Q_{a1}K_{a1}^\mathrm{T}}{\sqrt{d{_h}}}+M)V_{a1})
\end{equation}

 where FFN represents Feed Forward Network, $Q_{a1}$, $K_{a1}$, $V_{a1}$ are the projection mapping of the feature $x_{a1}$, $\mathrm{T}$ means transposition operation, $d_h$ is a scaled factor. 




 For the weak emotional correlation windows, the prior knowledge learned from the former block indicates that they have a high probability to be redundant for emotion recognition, so the features of the tokens located in them are multiplied by a weight $\lambda(\le1)$, while those in strong emotional correlation windows are multiplied by 1. The output is defined as $x_{a2} \in \mathbb{R}^{T \times D}$.

Temporal importance of each token within  window is calculated by IC module. Calculation results of all windows are then concatenated together into a sequence along the chronological order, which is noted as $impt_{a2} \in \mathbb{R}^T$. $x_{a2}$ and $impt_{a2}$ are passed to Dynamic Global Window Transformer module for further operation.
    
    
\subsubsection{Dynamic Global Window Transformer Module}

Dynamic Global Window Transformer (DGWT) module takes a holistic approach to remeasure the importance relationship between windows after DLWT.
In detail, each window firstly generates a token through \textbf{Window Weighted Sum} operation, which is defined as:


\begin{equation}
    wt_{k} = \sum_{p={b_{w_{k}}}}^{e_{w_{k}}}impt_{a2{_p}} \times  x_{a2_{p}}
\end{equation}
where $p$ is the index of the token. $wt \in \mathbb{R}^{(A+B) \times D}$ is the sequence of window tokens.

Then the sequence $wt$ is passed through a transformer encoder for global interaction, which is defined as:


\begin{equation}
    DGWT(wt) = FFN(Softmax(\frac{Q_{wt}K_{wt}^\mathrm{T}}{\sqrt{d_h}})V_{wt})
\end{equation}
where $Q_{wt}$, $K_{wt}$, $V_{wt}$ are the projection mapping of $wt$. 

Next, each window token is upsampled to the same length of the corresponding window by copying the vectors of the window. Then these tokens are concatenated together into a sequence, which is noted as $x_{a3} \in \mathbb{R}^{T \times D}$. The output of a DWFormer block $x_{(a+1)1}$ is the summation of $x_{a2}$ and $x_{a3}$ so that each token obtains both local and global information. 

The importance scores between windows $impt_{a3} \in \mathbb{R}^{A+B}$ are calculated by IC. Through a DWFormer block, the importance of each token $impt_{(a+1)1}$ is remeasured by:

\begin{equation}
    \resizebox{.91\linewidth}{!}{$impt_{(a+1)1} = Softmax(impt_{a2} \times Upsampling(impt_{a3}))$}
\end{equation}
the upsampling operation is the same as mentioned above.

In the next DWFormer block, $x_{(a+1)1}$ is split into windows based on $impt_{(a+1)1}$. 
Finally, the emotion classification is performed by applying the temporal average pooling layer on the output feature $x_{(N+1)1}$ of the $N$th DWFormer block, followed by a multi layer perception classifier. 

\section{Experiment}
\label{sec:pagestyle}



\subsection{Experiment Setup}

We evaluate DWFormer on IEMOCAP and MELD datasets. On IEMOCAP dataset, DWFormer is evaluated using 5-fold leave-one-section-out cross validation. 4 emotions (happy $\&$excited, angry, sad and neutral) are selected for classification. Weighted Accuracy (WA) and Unweighted Accuracy (UA) are the measuring metrics. On MELD corpus which contains 7 emotions (anger, disgust, fear, joy, neutral, sadness, surprise), the Weighted F1 (WF1) score is reported on test set.

 The output feature of the $12$th transformer encoder layer of Pre-trained WavLM-Large\cite{chen2022wavlm} model is used as the audio feature. 
 The number of DWFormer blocks for IEMOCAP is 3 and for MELD is 2. The number of heads is 8. The activation function is ReLU. The number of batchsize is 32. The learning rate is initialized to be 3e-4 for IEMOCAP, while 5e-4 for MELD. The value of $\lambda$ is 0.85. We employ an SGD optimizer for 120 epochs using a cosine decay learning rate scheduler with cosine warm-up scheduler. The optimization function is Cross Entropy Loss.


\subsection{Comparison to Baseline Networks}

The vanilla transformer, together with fixed window transformer which splits input feature into equal-length windows and applies self-attention within each window, are selected as the baseline networks. The parameters of baseline networks are the same as DWFormer. The window length of fixed window transformer is the same as the average length of the windows in the DWFormer. To verify the validity of the modules from DWFormer, ablation experiments are also conducted. 


Results in Table 1 demonstrate that DWFormer outperforms the Vanilla transformer and the fixed window transformer on both IEMOCAP and MELD datasets. 
Meanwhile, removing either the DLWT or the DGWT modules from DWFormer causes a significant decrease. 
\begin{table}[ht]
    \centering
    \scalebox{0.9}{
    \begin{tabular}{c|c|c|c}
    \hline
    \multirow{2}*{\textbf{Model}} & \multicolumn{2}{c|}{\textbf{IEMOCAP}}& \textbf{MELD} \\
    \cline{2-4}
    & \textbf{WA(\%)} & \textbf{UA(\%)} & \textbf{WF1(\%)} \\
    \hline
    Vanilla Transformer & 70.7 & 71.9 & 47.1 \\
    \hline
    Fixed Window Transformer & 71.2 & 72.3 & 47.6 \\
    \hline
    DWFormer (Ours) w/o DLWT & 71.5 & 72.4 & 47.8 \\
    \hline
    DWFormer (Ours) w/o DGWT & 71.5 & 72.7 & 47.7 \\
    \hline
    DWFormer (Ours) & \textbf{72.3} & \textbf{73.9} & \textbf{48.5} \\
    \hline
    \end{tabular}}
    \caption{Comparison results to baseline networks.}
    \label{tab:my_label}
\end{table}

\subsection{Comparison to conventional research\& Visualization}
Since our model employs the local-global architecture, we have conducted the comparison experiment with the conventional studies. \cite{liu2022atda} is open-source, so we first reproduce the results of \cite{liu2022atda} to ensure the correctness of the codes, and then we test the model under our experimental settings (\textbf{Exp 1}). Since the codes of \cite{liu2020speech, liu2020temporal} are not publicly available, we test our model under the experimental settings described in their papers(\textbf{Exp 2}: randomly split to 80\% training set and 20\% testing set). Experimental results are shown in Table 2, which demonstrate the superiority of our model compared with the other conventional studies.


\begin{table}[ht]
    \centering
    \scalebox{0.75}{
    \begin{tabular}{c|c|c|c|c}
    \hline
    \multirow{2}*{\textbf{Experimental Setting}} & \multirow{2}*{\textbf{Model}} & \multicolumn{2}{c|}{\textbf{IEMOCAP}}& \textbf{MELD} \\
    \cline{3-5}
     & & \textbf{WA(\%)} & \textbf{UA(\%)} & \textbf{WF1(\%)} \\
    \hline
    \multirow{2}*{Exp 1} & \cite{liu2022atda} & 59.6 & 60.5 & 39.8 \\
     \cline{2-5}& DWFormer (Ours) & \textbf{72.3} & \textbf{73.9} & \textbf{48.5} \\
    \hline
    \multirow{3}*{Exp 2}& \cite{liu2020speech}  & 69.4 & 70.2 & - \\
    \cline{2-5}& \cite{liu2020temporal}  & 70.3 & 70.8 & - \\
    \cline{2-5}& DWFormer (Ours) & \textbf{76.3} & \textbf{77.2} & - \\
    \hline
    \end{tabular}}
    \caption{Comparison results to the other conventional studies.}
    \label{tab:my_label}
\end{table}
In addition, Visualization results are shown in Fig. 5.
As shown in Fig. 5, Vanilla Transformer, Fixed Window Transformer and ATDA are not as good as ours in locating important temporal information.



\begin{figure}[t]
    \centering
    \includegraphics[width = \linewidth]{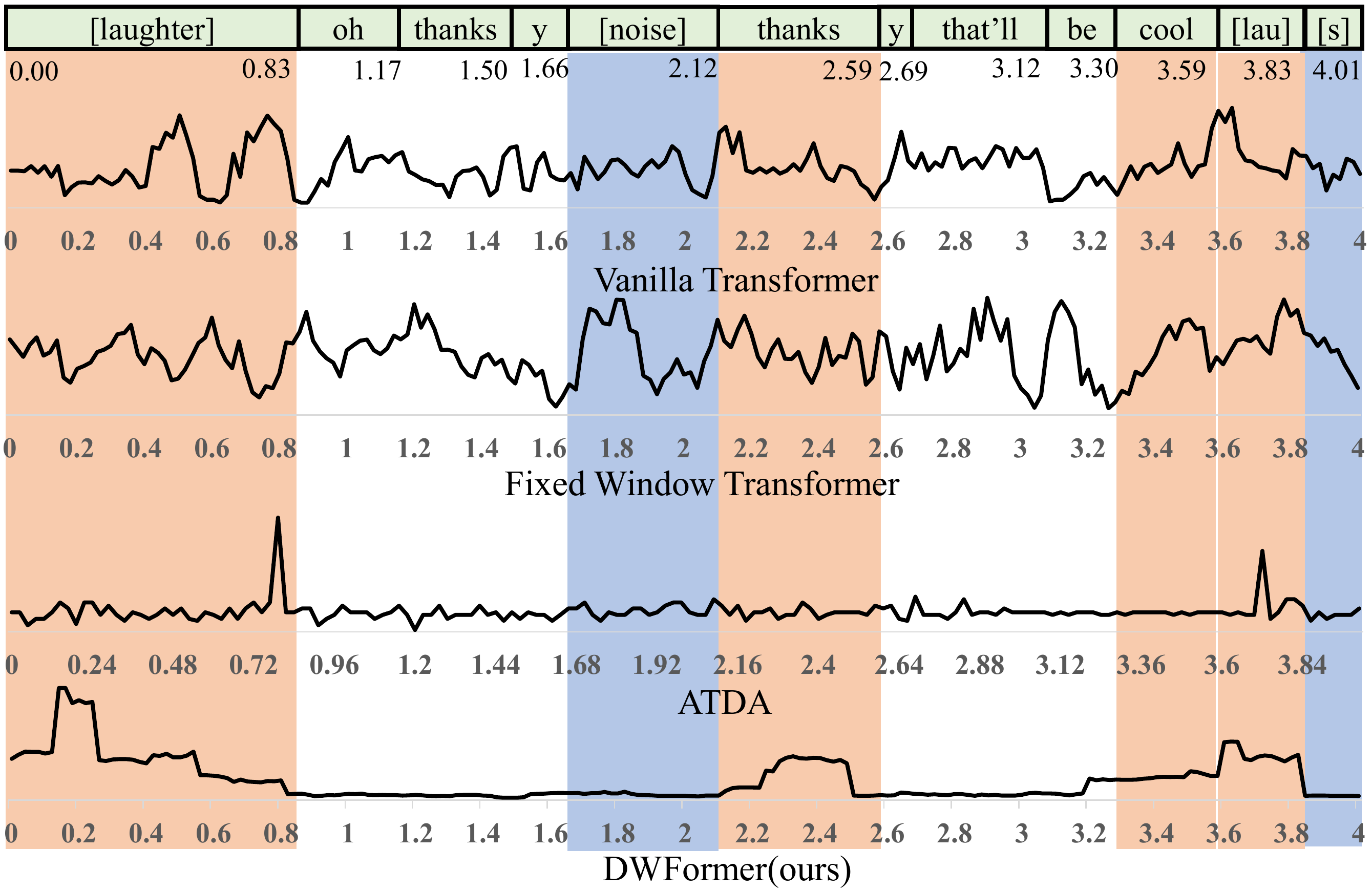}
    \caption{Visualization results of vanilla transformer, fixed window transformer, ATDA and DWFormer are shown. Horizontal axis indicates the chronological order and vertical axis is of importance score. [s] means silence, [lau] means laughter, y means 'yes'. Areas that are required to focus on are indicated by yellow borders, such as laughter, accent ('thanks'), positive semantic('cool'). Blue borders represent the area that should not be attended to, such as noise and silence.}
    \label{fig:my_label3}
\end{figure}
\subsection{Comparison to Previous State-of-the-art Methods}
Table 3 shows the comparison results between previous state-of-the-art methods and DWFormer. Experimental results prove that our method outperforms previous state-of-the-art methods. 
\begin{table}[ht]
    \centering
    \scalebox{0.75}{
    \begin{tabular}{c|c|c|c|c}
    \hline
    \textbf{Dataset} & \textbf{Model} & \textbf{WA(\%)} & \textbf{UA(\%)} & \textbf{WF1(\%)} \\
    \hline
     \multirow{4}*{IEMOCAP}& [Chen et al., 2022]\cite{chen22_interspeech} & 62.9 & 64.5 & -\\
    \cline{2-5} & [Li et al. 2022]\cite{li22v_interspeech} & 68.0 & 68.2 & -\\ 
    \cline{2-5} & [Zou et al., 2022]\cite{9747095} & 69.8 & 71.1 & - \\
    \cline{2-5} & DWFormer(Ours)    & \textbf{72.3} & \textbf{73.9} & - \\
    \hline
    \multirow{4}*{MELD} & [Chudasama et al., 2022]\cite{9747859} & - & - & 35.8\\
    \cline{2-5} & [Chen et al., 2022]\cite{chen22_interspeech} & - & - & 41.9\\
    \cline{2-5} & [Lian et al., 2022]\cite{9674867} & - & - & 45.2\\
    \cline{2-5} & DWFormer (Ours) & - & - & \textbf{48.5} \\
    \hline
    \end{tabular}}
    \caption{Comparison results to previous state-of-the-art methods.}
    \label{tab:my_label}
\end{table}

\section{Conclusions}
We propose a new transformer-based framework, DWFormer, which aims at capturing important temporal regions at variable scales within and across samples in SER field.
We empirically demonstrate that DWFormer outperforms the previous state-of-the-art methods. Ablation study proves the effectiveness of DLWT and DGWT modules. With the ability to locate important information, we plan to apply DWFormer in the pathological speech recognition field to assist researchers in understanding the impact of disease on pronunciation.

\section{Acknowledgement}
The work is supported in part by the Natural Science Foundation of Guangdong Province 2022A1515011588; the National Key R\&D Program of China (2022YFB4500600); the Science and Technology Project of Guangzhou 202103010002; the Science and Technology Project of Guangdong Guangdong 2022B0101010003; the National Natural Science Foundation of China under Grant U1801262; Guangdong  Provincial Key Laboratory of  Human Digital Twin (2022B1212010004).

\bibliographystyle{IEEEbib}
\bibliography{reference}

\end{document}